\documentclass[11pt]{amsart}
\usepackage{hyperref}
\theoremstyle{definition}

\theoremstyle{remark}

\numberwithin{equation}{section}

\def \dd{{\rm d}}
\begin{document}

\title[]{The Electrostatics of Einstein's Unified Field Theory}%
\author{Salvatore Antoci}%
\address{Dipartimento di Fisica ``A. Volta'' and CNR, Pavia, Italy}%
\email{Antoci@fisicavolta.unipv.it}%
\author{Dierck-Ekkehard  Liebscher}%
\address{Astrophysikalisches Institut Potsdam, Potsdam, Germany}%
\email{deliebscher@aip.de}%
\author{Luigi Mihich}%
\address{Dipartimento di Fisica ``A. Volta'', Pavia, Italy}%
\email{Mihich@fisicavolta.unipv.it}%

\begin{abstract}
When sources are added at their right-hand sides, and $g_{(ik)}$ is a priori
assumed to be the metric, the equations of Einstein's Hermitian
theory of relativity were shown to allow for an exact solution
that describes the general electrostatic field of $n$ point charges.
Moreover, the injunction of spherical symmetry of $g_{(ik)}$
in the infinitesimal neighbourhood of each of the charges was
proved to yield the equilibrium conditions of the $n$ charges in
keeping with ordinary electrostatics. The tensor $g_{(ik)}$,
however, cannot be the metric of the theory, since it enters
neither the eikonal equation nor the equation of motion
of uncharged test particles. A physically correct metric that
rules both the behaviour of wave fronts and of uncharged matter
is the one indicated by H\'ely.\par
In the present paper it is shown how the electrostatic solution
predicts the structure of the $n$ charged particles and their
mutual positions of electrostatic equilibrium when H\'ely's
physically correct metric is adopted.
\end{abstract}

\maketitle
\section{Introduction}
The Hermitian theory of relativity \cite{Einstein1948} is nearly
forgotten by the theoreticians of the present time; the few, who
still have some remembrance of the efforts done both by Einstein
\cite{Einstein1925, ES1946, Einstein1948, EK1955} and by Schr\"odinger
\cite{Schroedinger1947, Schroedinger1948, Schroedinger1951}
to find a generalization of the theory of 1915
that could encompass both gravitation and electromagnetism in a unique
geometrical structure, consider the matter as a subject
of purely historical interest.\par
One may observe, however, that no conclusive evidence was ever
brought against the theory, either through the exact solutions
\cite{Papapetrou1948, Wyman1950} or through
approximate calculations \cite{Callaway1953}. Nevertheless,
since no cogent identification of the geometrical objects
of the theory with physical entities had been achieved, the
interest aroused by the theory at the time of its appearance simply faded
away with the lapse of the years.\par
    A persistent prejudice, that has presumably undermined
the proper understanding of the theory, has been the
initial, tentative adoption, as metric tensor, of the symmetric
part $g_{(ik)}$ of the fundamental tensor. Not everybody has
incurred in such a prejudice; Lichnerowicz, for instance
\cite{Lichnerowicz1954, Lichnerowicz1955}, rightly based his choice of the metric
on the eikonal equation of the theory; however his argument
can determine that choice only up to an unknown
conformal factor. By challenging Einstein's and Schr\"odinger's
conviction, that the theory was a complete one, hence no source
terms were needed at the right-hand sides of the field
equations, H\'ely \cite{Hely1954a, Hely1954b} had succeeded in giving a
physically meaningful form to the conservation identities. He had
obtained this result by adopting as metric a particular tensor $s_{ik}$
that belongs to the class of conformally
related metrics allowed for by the argument of Lichnerowicz.
However, H\'ely's achievement went unnoticed,
apart from a few exceptions, e.g. \cite{Treder1957}.\par
When the class of exact solutions depending on three coordinates and
reported in \cite{Antoci1987a} was
found, the solutions were analysed by assuming $g_{(ik)}$ as metric.
It is mandatory to reconsider all these exact solutions by
adopting H\'ely's choice. We begin here, by
reconsidering what already seemed, with $g_{(ik)}$ as metric,
to be a sort of general electrostatic solution. We show that
it fully deserves such a title when the physically correct metric
is adopted.

\section{The field equations.}

A given geometric quantity \cite{Schouten1954} will be called hereafter
Hermitian with respect to the indices $i$ and $k$, both either covariant
or contravariant, if the part of the quantity that is symmetric with respect
to $i$ and $k$ is real, while the part that is antisymmetric is purely imaginary.
By extending into the complex domain the symmetry postulates of
general relativity, let us consider the Hermitian fundamental form
$g_{ik}=g_{(ik)}+g_{[ik]}$ and the affine connection
$\Gamma^i_{kl}=\Gamma^i_{(kl)}+\Gamma^i_{[kl]}$,
Hermitian with respect to the lower indices; both entities depend
on the real coordinates $x^i$, with $i$ running from 1 to 4. We
define also the Hermitian contravariant tensor $g^{ik}$ by the relation
\begin{equation}\label{2.1}
g^{il}g_{kl}=\delta^i_k,
\end{equation}
and the contravariant tensor density
$\mathbf{ g}^{ik}=(-g)^{1/2}g^{ik}$, where $g\equiv\text{det} (g_{ik})$
is a real quantity. Then the field equations of Einstein's unified field
theory in the complex Hermitian form \cite{Einstein1948} read
\begin{eqnarray}\label{2.2}
g_{ik,l}-g_{nk}\Gamma^n_{il}-g_{in}\Gamma^n_{lk}=0,\\\label{2.3}
\mathbf{ g}^{[is]}_{~~,s}=0,\\\label{2.4}
R_{(ik)}(\Gamma)=0,\\\label{2.5}
R_{[ik],l}(\Gamma)+R_{[kl],i}(\Gamma)+R_{[li],k}(\Gamma)=0;
\end{eqnarray}
$R_{ik}(\Gamma)$ is the Ricci tensor
\begin{equation}\label{2.6}
R_{ik}(\Gamma)=\Gamma^a_{ik,a}-\Gamma^a_{ia,k}
-\Gamma^a_{ib}\Gamma^b_{ak}+\Gamma^a_{ik}\Gamma^b_{ab}.
\end{equation}

\section{The general electrostatic solution.}

 When referred to the coordinates $x^1=x$, $x^2=y$, $x^3=z$,
$x^4=t$, the fundamental form $g_{ik}$ of the general
electrostatic solution \cite{Antoci1987b} of Einstein's unified field
theory in the Hermitian version reads
\begin{equation}\label{3.1}
g_{ik}=\left(\begin{array}{rrrr}
 -1 &  0 &  0 & a \\
  0 & -1 &  0 & b \\
  0 &  0 & -1 & c \\
 -a & -b & -c & d
\end{array}\right),
\end{equation}
with
\begin{equation}\label{3.2}
d=1+a^2+b^2+c^2
\end{equation}
and
\begin{equation}\label{3.3}
a=i\chi_{,x}, \ b=i\chi_{,y}, \ c=i\chi_{,z}, \ \
\chi_{,xx}+\chi_{,yy}+\chi_{,zz}=0.
\end{equation}

This particular solution of Einstein's unified field theory
belongs to a class of solutions depending on three coordinates
\cite{Antoci1987a}, outlined in Appendix \ref{A}.\par
In the ``Bildraum'' $x$, $y$, $z$, $t$ the imaginary part $g_{[ik]}$ of this
solution just looks like the general electrostatic solution of Maxwell's theory,
because the ``potential'' $\chi$ must obey the Laplace equation.
This is not, however, just a sort of ``Bildraum'' deception, for both
equations $\mathbf{ g}^{[is]}_{~~,s}=0$ and $g_{[[ik],l]}=0$ happen to be
satisfied.\par
If we allow for singularities at the right-hand
side of the field equations, the
electrostatic field due to $n$ point charges $h_q$, located at
$x=x_q$, $y=y_q$, $z=z_q$, can be built by taking
\begin{equation}\label{3.4}
\chi=-\sum_{q=1}^n\frac{h_q}{p_q},
\end{equation}
where
\begin{equation}\label{3.5}
p_q=[(x-x_q)^2+(y-y_q)^2+(z-z_q)^2]^{1/2}.
\end{equation}

We expect that, if we use singularities to represent charges and currents,
Einstein's Hermitian extension of the theory of general relativity
should give more information than Max\-well's equations do: it should predict
also the equations of motion of charges and currents, i.e., in the
case of the general electrostatic solution, the law for the electrostatic
equilibrium of the charges. We have agreed to represent charges
by singularities, and previous experience with the problem of
motion in general relativity has shown that the
behaviour of the field in an infinitesimal neighbourhood of the
singularities that represent the masses needs to be restricted in order to get
the equations of motion. In their ground-breaking paper of 1949,
Einstein and Infeld \cite{EI1949} did show that the equations of
motion of $n$ massive particles could be obtained by approximation
methods from the vacuum field equations of general relativity
alone if the metric was required to be spherically symmetric in the
infinitesimal neighbourhood of each of the $n$ particles.\par
We aim at imposing the same condition to the exact electrostatic solution
for which, according to (\ref{3.4}), $\chi$ looks in the ``Bildraum''
like the field of $n$ point charges, but of course imposing
the spherical symmetry can only be done if we know what symmetric
tensor represents the metric in our theory. We assume at first
that the metric be given by $g_{(ik)}$.\par
    Let Greek indices label henceforth
the coordinates $x^1=x$, $x^2=y$, $x^3=z$; then, according
to (\ref{3.1}), $g_{(\mu\nu)}$ acts as spatial metric. It is
just the Euclidean one, hence it is always spherically symmetric in
the infinitesimal neighbourhood of each charge, whatever the mutual
positions of the $n$ pointlike charges $h_q$ may be. This is not the case,
however, for the only nonvanishing component of $g_{(ik)}$ left,
i.e. $g_{44}=d$. With our coordinates it reads:
\begin{eqnarray}\label{3.6}
d=1-\sum^n_{q=1}\frac{h_q^2}{p_q^4}\\\nonumber
-\sum^{n (q\neq q')}_{q,q'=1}
h_qh_{q'}\frac{(x-x_q)(x-x_{q'})+(y-y_q)(y-y_{q'})
+(z-z_q)(z-z_{q'})}{p_q^3p_{q'}^3}.\nonumber
\end{eqnarray}
Let us examine $d$ in the infinitesimal neighbourhood of, say, the
$q$th charge. Due to the cross terms in the second line of (\ref{3.6}),
in general $d$ will not tend to a spherically symmetric behaviour
when the $q$th charge is approached; it will
do so only provided that the other charges are of such strengths
$h_{q'}$ and at such spatial positions $x_{q'}$, $y_{q'}$, $z_{q'}$
that:
\begin{equation}\label{3.7}
\sum^n_{q'\neq q}h_{q'}\frac{x_q-x_{q'}}{r_{qq'}^3}
=\sum^n_{q'\neq q}h_{q'}\frac{y_q-y_{q'}}{r_{qq'}^3}
=\sum^n_{q'\neq q}h_{q'}\frac{z_q-z_{q'}}{r_{qq'}^3}
=0,
\end{equation}
where $r_{qq'}$ is the Euclidean distance between the two charges
$q$ and $q'$, the one measured by the components $g_{(\mu\nu)}$
of the metric $g_{(ik)}$ chosen above. Equations (\ref{3.1}) and (\ref{3.7})
assert that, if $g_{(ik)}$ is the metric tensor of Einstein's unified
field theory, the equilibrium conditions for $n$ point electric charges
at rest predicted by an exact solution of that theory are just the same
as the ones occurring in the electrostatics of Coulomb.

\section{Choosing the metric of Einstein's unified field theory.}
In the previous Section we have tentatively chosen $g_{(ik)}$ to
be the metric of the theory in order to obtain a preliminary
reading of its possible content, but we have provided no
theoretical argument for this choice. The very fact that, if
$g_{(ik)}$ is taken as metric, then Einstein's unified field
theory contains an exact replica of Coulomb's electrostatics
is not an argument of a general character, and may well be
misleading. Our choice must stand on general theoretical
arguments \cite{Liebscher1985}, dealing with the
dynamics of waves and particles
predicted by the theory. One such argument was provided for Einstein's
unified field theory by Lichnerowicz \cite{Lichnerowicz1954, Lichnerowicz1955}.
As a consequence of his study of the Cauchy problem in
Einstein's unified field theory, he concluded that
the metric $l^{ik}$ appearing in the eikonal equation
\begin{equation}\label{4.1}
l^{ik}\partial_if\partial_kf=0
\end{equation}
for the wave surfaces of the theory had to be
\begin{equation}\label{4.2}
l^{ik}=g^{(ik)},
\end{equation}
or, one must add, any metric conformally related to $g^{(ik)}$.
Therefore, the argument by Lichnerowicz excludes the choice
of $g_{(ik)}$ as the metric of the theory.\par
If we adhere to Einstein's and Schr\"odinger's original
idea that, since the theory did represent the
completion of the theory of 1915, there was no need
to append sources at the right-hand side of equations
(\ref{2.2})-(\ref{2.5}), no further argument seems available for
further restricting the choice of the metric because,
during the decades elapsed since the theory was first proposed,
no tenable identification of physical entities in exact,
everywhere regular solutions was achieved, hence no determination
of the motion of matter predicted by the theory can be obtained
from the contracted Bianchi identities.\par This is no longer
true if, in keeping with what appears from the exact solutions,
one accepts the idea of appending sources
to the field equations of Einstein's unified field theory, just
like one does in general relativity, as first proposed in the
seminal work of H\'ely \cite{Hely1954a, Hely1954b}, outlined in Appendix \ref{B}.
But the analysis of the exact solutions \cite{Antoci1987b} shows that there is merit
if, by extending the approach of H\'ely, while retaining his choice of
the metric, sources are appended to all the original
field equations (\ref{2.2})-(\ref{2.5}). The way for doing so has
already been found \cite{Antoci1991}. In order to preserve
the Hermitian symmetry of the equations also when
the four-current density
$\mathbf{ j}^i=\frac{1}{4\pi}\mathbf{ g}^{[is]}_{~~,s}$
is nonvanishing, it is necessary to substitute the symmetrised
Ricci tensor of Borchsenius \cite{Borchsenius1978} for the plain
Ricci tensor (\ref{2.6}). However, since the symmetrised Ricci tensor
reduces to the plain one wherever  $\mathbf{ j}^i=0$, the way for
adding sources of \cite{Antoci1991} allows retaining the
original equations (\ref{2.2})-(\ref{2.5}) in vacuo.
It is reported in Appendix \ref{C}; it entails that the metric
must be the one chosen by H\'ely, defined in Appendix \ref{B} by the
equation
\begin{equation}\label{4.3}
s^{ik}=\frac{\sqrt{-g}}{\sqrt{-s}}g^{(ik)},
\end{equation}
where $s^{ik}$ is the inverse of H\'ely's metric $s_{ik}$, and
$s\equiv\text{det}(s_{ik})$.
The dynamical equations (\ref{C22}) for
charged matter obtained in Appendix \ref{C} from the contracted Bianchi
identities thus happen to require as metric just one among the conformally related
metrics that, according to Lichnerowicz \cite{Lichnerowicz1954, Lichnerowicz1955}, must
enter the eikonal equation (\ref{4.1}). Therefore, if H\'ely's
metric $s_{ik}$ is adopted in Einstein's unified field theory with sources,
the dynamics of both waves and particles is ruled by one and
the same metric. This was not the case with the former, tentative choice
of $g_{(ik)}$ as metric tensor.

\section{The general electrostatic solution when
the metric is $s_{ik}$.}

Let us reassess the general electrostatic solution
(\ref{3.1})-(\ref{3.3}) under the assumption that the
metric is the tensor $s_{ik}$ defined by (\ref{4.3}).
Due to (\ref{3.3}), for that solution $s_{ik}$ reads
\begin{eqnarray}\label{5.1}
s_{ik}=\sqrt{d}
\left(\begin{array}{rrrr}
 -1 &  0 &  0 & 0 \\
  0 & -1 &  0 & 0 \\
  0 &  0 & -1 & 0 \\
  0 &  0 &  0 & 1
\end{array}\right)\\\nonumber
-\frac{1}{\sqrt{d}}
\left(\begin{array}{cccc}
 \chi_{,x}\chi_{,x} & \chi_{,x}\chi_{,y} &  \chi_{,x}\chi_{,z} & 0 \\
 \chi_{,x}\chi_{,y} & \chi_{,y}\chi_{,y} &  \chi_{,y}\chi_{,z} & 0 \\
 \chi_{,x}\chi_{,z} & \chi_{,y}\chi_{,z} &  \chi_{,z}\chi_{,z} & 0 \\
 0 & 0 & 0 & 0
\end{array}\right),
\end{eqnarray}
hence the square of the line element can be written as
\begin{equation}\label{5.2}
\dd s^2=s_{ik}\dd x^{i}\dd x^{k}
=-\sqrt{d}\left(\dd x^2+\dd y^2+\dd z^2-\dd t^2\right)
-\frac{1}{\sqrt{d}}(\dd\chi)^2.
\end{equation}
If only one point charge $h$ is present in the ``Bildraum''
$x$, $y$, $z$, say, at the origin of the coordinates,
the ``potential'' $\chi$ is
\begin{equation}\label{5.3}
\chi=-\frac{h}{(x^2+y^2+z^2)^{1/2}}.
\end{equation}
In this case $s_{ik}$ is spherically symmetric, like
$g_{(ik)}$, but an essential difference in behaviour appears.
In fact, the manifold on which $g_{(ik)}$ is defined extends
to the full representative space $x$, $y$, $z$, although
$g_{(ik)}$ becomes a negative definite metric inside the spherical
surface where $d=0$. Let us remark in passing that the quantity $d$
appears as the $g_{44}$ component of the fundamental tensor (\ref{3.1}),
but has an invariant character, since it can be written
as $d=1-\frac12g^{[ik]}g_{[ik]}$. When the metric is given by $s_{ik}$, the
surface $d=0$ instead constitutes the inner border of the manifold,
since the square root of $d$ occurs in (\ref{5.2}).
With this metric, the charge can no longer reside at the origin
of the coordinates; it must stay on the sphere $d=0$, whose
squared coordinate radius is $r_h^2=x^2+y^2+z^2=|h|$. But is the surface $d=0$ of
the representative space a surface also in the metric sense?
Let us consider the spheres centered
at the origin, whose coordinate radius is larger than $r_h$.
They are equipotential surfaces; we have in fact
\begin{equation}\label{5.4}
\dd \chi=0
\end{equation}
for any infinitesimal, spatial displacement $\dd x^{\mu}$ constrained to
occur on one of these spheres. Therefore the last term of
(\ref{5.2}) is always zero on each sphere, and the residual square of the spatial
interval tends to a vanishing value when measured
on a sphere whose coordinate radius approaches $r_h$,
since it contains the factor $\sqrt{d}$. In this
sense, one can conclude that the surface $d=0$ is in fact a point,
hence the charge associated with the ``potential''
(\ref{5.3}) can be deemed pointlike also when its spatial dimension
is measured by the metric $s_{ik}$.\par
If more than one point charge is present in the ``Bildraum'', and
the ``potential'' $\chi$ is no longer given by (\ref{5.3}), but by
(\ref{3.4}) and (\ref{3.5}), we may have $n$ closed surfaces
on which $d=0$ but, whatever the positions of the point charges in
the representative space may be, they will not be equipotential
surfaces. Therefore, since $\dd \chi\ne 0$ when $d=0$, the last term of the
squared interval (\ref{5.2}) will be infinite, hence there will be no room
for point charges and for their equilibrium conditions in the
metric sense. At variance with what occurs when the metric is
assumed to be $g_{(ik)}$, introducing point charges in the ``Bildraum'' is
not the right way for getting them eventually.\par
The very form of (\ref{5.2}) suggests however an alternative
choice: if $\chi$ is just the potential that the
electrostatics of Coulomb attributes to $n$ charged conductors, and their
closed surfaces are so chosen that $d>0$ on each of them, the last term
of (\ref{5.2}) will always be zero for displacements occurring on
these surfaces. We can then imagine altering the shape and the relative
positions of these charged surfaces in the representative
space by always retaining their equipotential quality, just as it
would spontaneously occur in an actual experiment done with charged conductors.
If we succeed in this way in getting $d=0$ on each of these
surfaces, according to (\ref{5.2}) the charges will be pointlike
in the metric sense, and they will be in their mutual positions of
equilibrium, since the metric $s_{ik}$ will be spherically
symmetric in an infinitesimal neighbourhood of each charged
point.\par
Solving such a mathematical problem is obviously
{\it extra vires} when the shapes and the mutual positions
of the charged surfaces in the representative space $x$, $y$, $z$
are allowed for full generality. However, the problem becomes much
simpler if, in the ``Bildraum'', the charged conducting surfaces
are spherical and mutually far apart. In this case $\chi$ will behave
just like the potential of $n$ charged spherical conductors
whose mutual separations are very large with respect to the
radii of the spheres, and will closely approximate, outside
the spheres, the potential (\ref{3.4}) of $n$ point charges $h_q$ residing
at the center of the spheres. If the centers are not in the
positions of equilibrium given by (\ref{3.7}), the mutual
inductions that render the conducting spheres equipotential,
thereby ensuring the vanishing of the last term of the interval
(\ref{5.2}) on each charged surface, will produce an inhomogeneous
surface charge density. Therefore, due to Coulomb's theorem, the value of $d$
on each surface will not be constant.\par
If instead the centers are in the positions given by (\ref{3.7}),
the mutual inductions will become very small since, in the representative space,
the radii of the spheres have been chosen to be very small
with respect to their mutual separations. Therefore the value of
$d$ will be nearly constant on each charged surface, i.e. the interval
will be nearly spherically symmetric in an infinitesimal neighbourhood
of the surface. By taking spheres with smaller
and smaller radii we can thus approach both the condition of spherical
symmetry and the condition $d=0$, that ensures the pointlike
character of the charges.\par
Of course we cannot attain an exact result in the sense of a
limit, because the radii of the spheres must remain finite
in the representative space, to avoid that $d$ become negative.
However, since the largest electric fields observed until now
do not seem to affect the metric properties of space in a
considerable way, we are ensured that the approximate conditions
of equilibrium and the approximate pointlike structure of the
charges obtained by the procedure outlined above are precise enough
when confronted with the most stringent empirical constraints.\par
We eventually remark that the condition that $\dd\chi$ be
vanishing on the surfaces where the charges reside is
equivalent to the invariant condition that the metric (\ref{5.1})
be conformally flat just on these surfaces.

\section{Conclusion}
When sources are allowed for at the right-hand sides
of the field equations  (\ref{2.2})-(\ref{2.5}) of
Einstein's Hermitian theory of relativity in the way
shown in Appendix \ref{C}, the conservation identities take
the physically expressive form (\ref{C22}) provided that
$s_{ik}$, defined by (\ref{4.3}), is chosen as metric.\par
The paradigmatic exact solution (\ref{3.1})-(\ref{3.3}) happens to fully
deserve the name of electrostatic solution with the metric $s_{ik}$ too.
The approximate pointlike structure of $n$ charges
and their approximate equilibrium conditions stem in fact from
the purely geometrical condition of spherical symmetry
in the neighbourhood of each charge also when the new,
physically correct metric is adopted.\par

\appendix
\section{Solutions depending on three coordinates}\label{A}
Let the real symmetric tensor $h_{ik}$ be the metric for a
solution to the field equations of general relativity, which
depends on the first three co-ordinates $x^{\lambda}$, not necessarily
all spatial in character, and for which
$h_{\lambda 4}=0$. We assume Greek indices to run
from 1 to 3, while Latin indices run from 1 to 4. We consider also an
antisymmetric purely imaginary tensor $a_{ik}$, which depends on
the first three co-ordinates, and we assume that its only nonvanishing
components are $a_{\mu 4}=-a_{4 \mu}$. Then we form the mixed tensor
\begin{equation}\label{A1}
\alpha_i^{~k}=a_{il}h^{lk}=-\alpha^k_{~i},
\end{equation}
where $h^{ik}$ is the inverse of $h_{ik}$, and we define the
Hermitian fundamental form $g_{ik}$ as follows:
\begin{eqnarray}\nonumber
g_{\lambda\mu}=h_{\lambda\mu},\\\label{A2}
g_{4\mu}=\alpha_4^{~\nu}h_{\nu\mu},\\\nonumber
g_{44}=h_{44}-\alpha_4^{~\mu}\alpha_4^{~\nu}h_{\mu\nu}.
\end{eqnarray}
When the three additional conditions
\begin{equation}\label{A3}
\alpha^4_{~\mu,\lambda}-\alpha^4_{~\lambda,\mu}=0
\end{equation}
are fulfilled, the affine connection $\Gamma^i_{kl}$ which solves
eqs. (\ref{2.2}) has the nonvanishing components
\begin{eqnarray}\label{A4}
\Gamma^{\lambda}_{(\mu\nu)}=\left\{^{~\lambda}_{\mu~\nu}\right\}_h,
\\\nonumber
\Gamma^{\lambda}_{[4\nu]}=\alpha^{~\lambda}_{4~,\nu}
-\left\{^{~4}_{4~\nu}\right\}_h\alpha^{~\lambda}_4
+\left\{^{~\lambda}_{\rho~\nu}\right\}_h\alpha^{~\varrho}_4,
\\\nonumber
\Gamma^4_{(4\nu)}=\left\{^{~4}_{4~\nu}\right\}_h,
\\\nonumber
\Gamma^{\lambda}_{44}=\left\{^{~\lambda}_{4~4}\right\}_h
-\alpha^{~\nu}_4\left(\Gamma^{\lambda}_{[4\nu]}
-\alpha^{~\lambda}_4\Gamma^4_{(4\nu)}\right);
\end{eqnarray}
we indicate with $\left\{^{~i}_{k~l}\right\}_h$ the Christoffel
connection built with $h_{ik}$. We form now the Ricci tensor (\ref{2.6}).
When eqs. (\ref{2.3}), i.e., in
our case, the single equation
\begin{equation}\label{A5}
(\sqrt{-h}~\alpha^{~\lambda}_4 h^{44})_{,\lambda}=0,
\end{equation}
and the additional conditions, expressed by eqs. (\ref{A3}), are
satisfied, the components of $R_{ik}(\Gamma)$ can be written as
\begin{eqnarray}\nonumber
R_{\lambda\mu}=H_{\lambda\mu},
\\\label{A6}
R_{4\mu}=\alpha^{~\nu}_4H_{\nu\mu}+\left(\alpha^{~\nu}_4
\left\{^{~4}_{4~\nu}\right\}_h\right)_{,\mu},
\\\nonumber
R_{44}=H_{44}-\alpha^{~\mu}_4\alpha^{~\nu}_4H_{\mu\nu},
\end{eqnarray}
where $H_{ik}$ is the Ricci tensor built with
$\left\{^{~i}_{k~l}\right\}_h$. $H_{ik}$ is zero when $h_{ik}$ is a
solution of the field equations of general relativity, as
supposed; therefore, when eqs. (\ref{A3}) and (\ref{A5}) hold, the
Ricci tensor, defined by eqs. (\ref{A6}), satisfies eqs. (\ref{2.4})
and (\ref{2.5}) of the Hermitian theory of relativity.\par
The task of solving equations (\ref{2.2})-(\ref{2.5}) reduces, under
the circumstances considered here, to the simpler task of solving
eqs. (\ref{A3}) and (\ref{A5}) for a given $h_{ik}$.\par
We eventually note that the method applies also
to Schr\"odin\-ger's purely affine theory \cite{Schroedinger1951}.

\section{H\'ely's proposal for the metric and for the sources}\label{B}
When equations (\ref{2.2}) and (\ref{2.3}) of Einstein's unified field theory
hold, the contracted Bianchi identities
take the form \cite{ Schroedinger1948}
\begin{equation}\label{B1}
\left[\sqrt{-g}\left(g^{ik}R_{il}+g^{ki}R_{li}\right)\right]_{,k}
=\sqrt{-g}g^{ik}R_{ik,l},
\end{equation}
or else
\begin{eqnarray}\label{B2}
\left(2\sqrt{-g}g^{(ik)}R_{(il)}\right)_{,k}
-\sqrt{-g}g^{(ik)}R_{(ik),l}\\\nonumber
=\sqrt{-g}g^{[ik]}\left(R_{[ik],l}+R_{[kl],i}+R_{[li],k}\right).
\end{eqnarray}
In his paper \cite{Hely1954a}, entitled {\it ``Sur la repr\'esentation d'Einstein du
champ unitaire''}, H\'ely introduces a symmetric tensor
$s^{ik}$ such that
\begin{equation}\label{B3}
\sqrt{-s}s^{ik}=\sqrt{-g}g^{(ik)},
\end{equation}
where $s$ is the determinant of the tensor $s_{ik}$, and
$s^{ik}s_{il}=\delta^k_l$. By introducing this tensor in the left-hand
side of (\ref{B2}) something unexpected occurs. H\'ely finds in
fact\footnote{In H\'ely's papers, $f_{ik}$ stands for our $s_{ik}$, and the real
nonsymmetric version of Einstein's unified field theory is considered.
His equations, however, remain formally unaffected when the complex
Hermitian version of the theory is considered instead.} that
\begin{eqnarray}\label{B4}
\left(2\sqrt{-g}g^{(ik)}R_{(il)}\right)_{,k}
-\sqrt{-g}g^{(ik)}R_{(ik),l}\\\nonumber
=\left(2\sqrt{-s}s^{ik}R_{(il)}\right)_{;k}
-\left(\sqrt{-s}s^{ik}R_{(ik)}\right)_{;l},
\end{eqnarray}
where the semicolon stands for the covariant differentiation
with respect to the Christoffel symbols built with $s_{ik}$.
Hence the contracted Bianchi identities of Einstein's nonRiemannian
extension of the vacuum general relativity of 1915 are shown
by H\'ely to admit a sort of Riemannian rewriting in terms of
the metric $s_{ik}$:
\begin{eqnarray}\label{B5}
\left(s^{ik}R_{(il)}
-\frac12\delta^k_l s^{pq}R_{(pq)}\right)_{;k}\\\nonumber
=\frac12\sqrt{\frac{g}{s}}g^{[ik]}\left(R_{[ik],l}+R_{[kl],i}+R_{[li],k}\right),
\end{eqnarray}
provided that equations (\ref{2.2}) and (\ref{2.3}) are satisfied.
However, since in Einstein's theory equations
(\ref{2.4}) and (\ref{2.5}) need to be satisfied too, the contracted
identities in the form (\ref{B5}) appear devoid of physical sense, because
both sides happen to be vanishing.\par
It has been already reminded that, according both to Einstein and to
Schr\"odinger, the equations (\ref{2.2})-(\ref{2.5}) did represent
the completion of the equations of 1915, hence no phenomenological sources
should be admitted at their right-hand sides. The form of (\ref{B5}) was
however so suggestive for H\'ely, that he dared challenging the opinion
mentioned above, and published a sequel \cite{Hely1954b}, entitled
{\it ``Sur une g\'en\'eralisation imm\'ediate des \'equations d'Einstein''},
to the previously recalled paper \cite{Hely1954a},
in which phenomenological sources are appended to the right-hand sides
of both (\ref{2.4}) and (\ref{2.5}). Thereby  a nontrivial content
is given to (\ref{B5}) as conservation identity, and a tentative physical
interpretation to the whole theory is advanced.\par
In \cite{Hely1954b} H\'ely proposes substituting the definitions:
\begin{eqnarray}\label{B6}
R_{(ik)}=T_{ik},\\\label{B7}
R_{[ik],l}+R_{[kl],i}+R_{[li],k}=4\pi J_{ikl},
\end{eqnarray}
where $T_{ik}$ is a symmetric tensor, while $J_{ikl}$ is a totally antisymmetric
one, for the equations (\ref{2.4}) and (\ref{2.5}).  By availing of
$s^{ik}$ for raising indices, (\ref{B5}) comes to read:
\begin{equation}\label{B8}
\left(T^k_l-\frac12\delta^k_lT^s_s\right)_{;k}
=2\pi\sqrt{\frac{g}{s}}g^{[ik]}J_{ikl}.
\end{equation}
Then the right-hand side of (\ref{B8}) describes the Lorentz force exerted by the
antisymmetric field $\sqrt{g/s}~g^{[ik]}$ on the conserved current $J_{ikl}$.
Pending the final verdict coming from the solutions to the field equations,
this electromagnetic interpretation appears to be a consistent one: by looking
at equations (\ref{B7}) and (\ref{2.3}) together,
H\'ely is entitled to assert that, in the generalization of Einstein's theory
proposed by him, $\sqrt{g/s}~g^{[ik]}$ and $R_{[ik]}$ are proportional
to the duals of the electromagnetic fields
$(\overrightarrow{E},\overrightarrow{B})$ and
$(\overrightarrow{D},\overrightarrow{H})$ respectively. Like
in the theory of 1915, the Lorentz force appears to be due to the
nonconservation of some stress-energy-momentum tensor associated
with the fields, whose expression is however much more complicated
that in the Maxwellian case, because the constitutive equation
linking $\sqrt{g/s}~g^{[ik]}$ and $R_{[ik]}$ is a very complicated, differential
relation, without counterpart in classical electromagnetism.

\section{Generalization of H\'ely's approach}\label{C}
The way of appending sources given in \cite{Antoci1991}
will be recalled here in full detail, in order to allow for the
comparison with H\'ely's original proposal.
On a four-dimensional, real manifold, let
$\mathbf{g}^{ik}$ be a Hermitian contravariant tensor density
\begin{equation}\label{C1}
\mathbf{g}^{ik}=\mathbf{g}^{(ik)}+\mathbf{g}^{[ik]}.
\end{equation}
We also endow the manifold with a general, complex affine connection
\begin{equation}\label{C2}
W^i_{kl}=W^i_{(kl)}+W^i_{[kl]};
\end{equation}
for the Riemann curvature tensor built with this connection:
\begin{equation}\label{C3}
R^i_{~klm}(W)=W^i_{kl,m}-W^i_{km,l}
-W^i_{al}W^a_{km}+W^i_{am}W^a_{kl},
\end{equation}
two distinct nonvanishing contractions \cite{Schroedinger1950},
namely $R_{ik}(W)=R^p_{~ikp}(W)$ and $A_{ik}(W)=R^p_{~pik}(W)$ do exist.
But also the transposed affine connection $\tilde{W}^i_{kl}=W^i_{lk}$ shall
be taken into account: from it, the Riemann curvature tensor
$R^i_{~klm}(\tilde{W})$ and its two contractions
$R_{ik}(\tilde{W})$ and $A_{ik}(\tilde{W})$ can be formed as well.
We aim at following the pattern of general relativity, in which the
Lagrangian density $\mathbf{g}^{ik}R_{ik}$ is considered, but now
any linear combination $\bar{R}_{ik}$ of the four above-mentioned
contractions can be envisaged. A good choice \cite{Borchsenius1978},
for reasons that will become apparent much later, is
\begin{equation}\label{C4}
\bar{R}_{ik}(W)=R_{ik}(W)+\frac{1}{2}A_{ik}(\tilde{W}).
\end{equation}
Let us try to endow the theory with sources in the form of a
nonsymmetric tensor $P_{ik}$ and of
a current density $\mathbf{j}^i$, coupled to
$\mathbf{g}^{ik}$ and to the vector $W_i=W^l_{[il]}$ respectively.
The Lagrangian density
\begin{equation}\label{C5}
\mathbf{L}=\mathbf{g}^{ik}\bar{R}_{ik}(W)
-8\pi\mathbf{g}^{ik}P_{ik}
+\frac{8\pi}{3}W_i\mathbf{j}^i
\end{equation}
is thus arrived at.  By performing independent variations of the
action $\int\mathbf{L}d\Omega$ with respect to $W^p_{qr}$ and to
$\mathbf{g}^{ik}$ with suitable boundary conditions we obtain the
field equations
\begin{eqnarray}\label{C6}
-\mathbf{g}^{qr}_{~,p}+\delta^r_p\mathbf{g}^{(sq)}_{~,s}
-\mathbf{g}^{sr}W^q_{sp}-\mathbf{g}^{qs}W^r_{ps}
+\delta^r_p\mathbf{g}^{st}W^q_{st}
+\mathbf{g}^{qr}W^t_{pt}\\\nonumber
=\frac{4\pi}{3}(\mathbf{j}^r\delta^q_p-\mathbf{j}^q\delta^r_p),
\end{eqnarray}
and
\begin{equation}\label{C7}
\bar{R}_{ik}(W)=8\pi P_{ik}.
\end{equation}
By contracting eq. (\ref{C6}) with respect to $q$ and $p$ one obtains
\begin{equation}\label{C8}
\mathbf{g}^{[is]}_{~~,s}={4\pi}\mathbf{j}^i,
\end{equation}
a promising outcome, but a problem too. In fact, the very
existence of (\ref{C8}) tells that, for given $\mathbf{j}^i$,
equation (\ref{C6}) cannot determine the affine connection
$W^i_{kl}$ uniquely in terms of $\mathbf{g}^{ik}$:
(\ref{C6}) is in fact invariant under the projective
transformation ${W'}^i_{kl}=W^i_{kl}+\delta^i_k\lambda_l$, where
$\lambda_l$ is an arbitrary vector field. Moreover eq. (\ref{C7})
is invariant under the transformation
\begin{equation}\label{C9}
{W'}^i_{kl}=W^i_{kl}+\delta^i_k\mu_{,l},
\end{equation}
where $\mu$ is an arbitrary scalar. Both equation (\ref{C8}) and the
invariance under (\ref{C9}) are reminiscent of electromagnetism
as we know it. We can write
\begin{equation}\label{C10}
W^i_{kl}=\Gamma^i_{kl}-\frac{2}{3}\delta^i_kW_l,
\end{equation}
where $\Gamma^i_{kl}$ is another affine connection, by definition
constrained to yield $\Gamma^l_{[il]}=0$. Then eq. (\ref{C6})
becomes
\begin{equation}\label{C11}
\mathbf{g}^{qr}_{~,p}+\mathbf{g}^{sr}\Gamma^q_{sp}+\mathbf{g}^{qs}\Gamma^r_{ps}
-\mathbf{g}^{qr}\Gamma^t_{(pt)}
=\frac{4\pi}{3}(\mathbf{j}^q\delta^r_p-\mathbf{j}^r\delta^q_p),
\end{equation}
and allows us to determine $\Gamma^i_{kl}$ uniquely, under very
general conditions, in terms of $\mathbf{g}^{ik}$. When eq.
(\ref{C10}) is substituted in eq. (\ref{C7}), the even and the
alternating part of the latter come to read:
\begin{eqnarray}\label{C12}
\bar{R}_{(ik)}(\Gamma)=8\pi P_{(ik)}\\\label{C13}
\bar{R}_{[ik]}(\Gamma)
=8\pi P_{[ik]}-\frac{1}{3}(W_{i,k}-W_{k,i})
\end{eqnarray}
respectively. An unsurmountable difficulty appears in equation (\ref{C13}).
In fact \cite{Schroedinger1950}, wherever a source term is nonvanishing, a field
equation loses its meaning, and reduces to a definition of some property of matter
in terms of geometrical entities; it is quite obvious that such a definition
must be unique. This necessary occurrence happens with eqs. (\ref{C11}),
(\ref{C8}) and (\ref{C12}), but it does not happen with eq. (\ref{C13}). This
equation only prescribes that $\bar{R}_{[ik]}(\Gamma)-8\pi P_{[ik]}$
is the curl of an arbitrary vector $W_i/3$; it is just equivalent to
the four equations
\begin{equation}\label{C14}
\bar{R}_{[[ik],l]}(\Gamma)=8\pi P_{[[ik],l]},
\end{equation}
hence it cannot specify $P_{[ik]}$ uniquely. Therefore we must dismiss the
redundant tensor $P_{[ik]}$, and assume henceforth that matter is defined by the
symmetric tensor $P_{(ik)}$, by the current density $\mathbf{j}^i$
and by the current
\begin{equation}\label{C15}
K_{ikl}=\frac{1}{8\pi}\bar{R}_{[[ik],l]};
\end{equation}
both $\mathbf{j}^i$ and $K_{ikl}$ are conserved quantities by
definition. The analogy with the general relativity of 1915, to which the
present theory formally reduces when $\mathbf{g}^{[ik]}=0$,
suggests rewriting  eq. (\ref{C12}) as
\begin{equation}\label{C16}
\bar{R}_{(ik)}(\Gamma)=8\pi(T_{ik}
-\frac{1}{2}s_{ik}s^{pq}T_{pq}),
\end{equation}
where $s_{ik}=s_{ki}$ is the still unchosen metric tensor of the
theory, $s^{il}s_{kl}=\delta^i_k$, and the symmetric tensor
$T_{ik}$ will be tentatively assumed to play the
r\^ole of energy tensor.\par
    Equations (\ref{C11}), (\ref{C8}), (\ref{C16}) and (\ref{C15})
reduce to the equations (\ref{2.2})-(\ref{2.5}) of Einstein's
unified field theory when sources are absent, since then
$\bar{R}_{ik}(\Gamma)$=${R}_{ik}(\Gamma)$; moreover they enjoy
the property of transposition invariance even when sources are
present. If $\mathbf{g}^{ik}$ is Hermitian, like it was assumed, $\Gamma^i_{kl}$,
as defined by equation (\ref{C11}), is Hermitian too, and the same
property is enjoyed also by $\bar{R}_{ik}(\Gamma)$.
Let these quantities represent a solution with the sources
$T_{ik}$, $\mathbf{j}^i$ and $K_{ikl}$. The transposed quantities
$\tilde{\mathbf{g}}^{ik}=\mathbf{g}^{ki}$,
$\tilde{\Gamma}^i_{kl}=\Gamma^i_{lk}$ and
$\bar{R}_{ik}(\tilde{\Gamma})$= $\bar{R}_{ki}(\Gamma)$ then
provide another solution, endowed with the sources
$\tilde{T}_{ik}=T_{ik}, \tilde{\mathbf{j}}^i=-\mathbf{j}^i$
and $\tilde{K}_{ikl}=-K_{ikl}$. Such a desirable property is a
consequence of the choice made for $\bar{R}_{ik}$. These equations
suggest interpreting Einstein's unified field theory with sources
as a gravoelectrodynamics in a polarizable continuum, allowing for
both electric and magnetic currents. The study of the conservation
identities confirms the idea and leads at the same time to the
determination of the metric tensor $s_{ik}$ that appears in
equation (\ref{C16}). One considers the
invariant integral
\begin{equation}\label{C17}
I=\int\left[\mathbf{g}^{ik}\bar{R}_{ik}(W)
+\frac{8\pi}{3}W_i\mathbf{j}^i\right]d\Omega.
\end{equation}
From it, when eq. (\ref{C6}) is assumed to hold, by means of an
infinitesimal coordinate transformation the four identities
\begin{eqnarray}\label{C18}
-(\mathbf{g}^{is}\bar{R}_{ik}(W)
+\mathbf{g}^{si}\bar{R}_{ki}(W))_{,s}
+\mathbf{g}^{pq}\bar{R}_{pq,k}(W)\\\nonumber
+\frac{8\pi}{3}\mathbf{j}^i(W_{i,k}-W_{k,i})=0
\end{eqnarray}
are obtained. This equation can be rewritten as
\begin{eqnarray}\label{C19}
-2(\mathbf{g}^{(is)}\bar{R}_{(ik)}(\Gamma))_{,s}
+\mathbf{g}^{(pq)}\bar{R}_{(pq),k}(\Gamma)\\\nonumber
=2\mathbf{g}^{[is]}_{~~,s}\bar{R}_{[ik]}(\Gamma)
+\mathbf{g}^{[is]}\bar{R}_{[[ik],s]}(\Gamma),
\end{eqnarray}
where the redundant variable $W^i_{kl}$ no longer appears. Let us
remind of eq. (\ref{C16}), and assume, like H\'ely did
\cite{Hely1954a, Hely1954b}, that the metric tensor is defined by the equation
\begin{equation}\label{C20}
\sqrt{-s}s^{ik}=\mathbf{g}^{(ik)},
\end{equation}
where $s=\det{(s_{ik})}$; we shall use henceforth $s^{ik}$ and
$s_{ik}$ to raise and lower indices, $\sqrt{-s}$ to produce
tensor densities out of tensors. We define then
\begin{equation}\label{C21}
\mathbf{T}^{ik}=\sqrt{-s}s^{ip}s^{kq}T_{pq},
\end{equation}
and the weak identities (\ref{C19}), when all the field equations
hold, will take the form
\begin{equation}\label{C22}
\mathbf{T}^{ls}_{;s}=\frac{1}{2}s^{lk}
(\mathbf{j}^i\bar{R}_{[ki]}(\Gamma)
+K_{iks}\mathbf{g}^{[si]}),
\end{equation}
where the semicolon indicates the covariant derivative done with
respect to the Christoffel connection
\begin{equation}\label{C23}
\left\{^{~i}_{k~l}\right\}
=\frac{1}{2}s^{im}(s_{mk,l}+s_{ml,k}-s_{kl,m})
\end{equation}
built with $s_{ik}$. Our earlier impression is confirmed by eq.
(\ref{C22}): the theory, built in terms of a non-Riemannian
geometry, entails a gravoelectro\-dynamics in a dynamically
polarized Riemannian spacetime, for which $s_{ik}$ is the metric.
Like in H\'ely's proposal \cite{Hely1954b} of 1954, the relationship between
electromagnetic inductions and fields is governed by the field equations in a
quite novel and subtle way, with respect to the one prevailing, say, in the
electromagnetic vacuum of the so-called Einstein-Maxwell theory; with aftersight,
one may well assert that this novelty has constituted, besides the choice of the metric,
another major stumbling block in the understanding of the theory.

\newpage

\bibliographystyle{amsplain}

\end{document}